\documentclass[twocolumn,showpacs,preprintnumbers,amsmath,amssymb]{revtex4}

\usepackage{graphicx}
\usepackage{multirow}

\begin{document}

\title{Zero-temperature magnetic response of small fullerene molecules at the classical and full quantum limit}

\author{N. P. Konstantinidis}
\affiliation{Department of Physics, King Fahd University of Petroleum and Minerals, Dhahran 31261, Saudi Arabia}

\date{\today}

\begin{abstract}
The ground-state magnetic response of fullerene molecules with up to 36 vertices is calculated, when spins classical or with magnitude $s=\frac{1}{2}$ are located on their vertices and interact according to the nearest-neighbor antiferromagnetic Heisenberg model. The frustrated topology, which originates in the pentagons of the fullerenes and is enhanced by their close proximity, leads to a significant number of classical magnetization and susceptibility discontinuities, something not expected for a model lacking magnetic anisotropy. This establishes the classical discontinuities as a generic feature of fullerene molecules irrespective of their symmetry. The largest number of discontinuities have the molecule with 26 sites, four of the magnetization and two of the susceptibility, and an isomer with 34 sites, which has three each. In addition, for several of the fullerenes the classical zero-field lowest energy configuration has finite magnetization, which is unexpected for antiferromagnetic interactions between an even number of spins and with each spin having the same number of nearest-neighbors. The molecules come in different symmetries and topologies and there are only a few patterns of magnetic behavior that can be detected from such a small sample of relatively small fullerenes.
Contrary to the classical case, in the full quantum limit $s=\frac{1}{2}$ there are no discontinuities for a subset of the molecules that was considered. This leaves the icosahedral symmetry fullerenes as the only ones known supporting ground-state magnetization discontinuities for $s=\frac{1}{2}$. It is also found that a molecule with 34 sites has a doubly-degenerate ground state when $s=\frac{1}{2}$.
\end{abstract}


\maketitle

\section{Introduction}
\label{sec:introduction}

The antiferromagnetic Heisenberg model (AHM) has been extensively investigated when describing the interactions of spins located at the vertices of low-dimensional frustrated topologies \cite{Auerbach98,Fazekas99,NPK02,NPK04,NPK13,Nakano14,Nakano14-1}. One such case are fullerene molecules, which are zero-dimensional \cite{Coffey92,NPK01,NPK05,NPK07,NPK09,NPK16,NPK17,Jimenez14}. Their low energy spectrum includes non-magnetic excitations inside the singlet-triplet gap, and their specific heat has a multi-peak structure as a function of temperature. Furthermore, they have a number of ground-state magnetization discontinuities in an external field at the classical and quantum level, even though such discontinuities are normally expected from Hamiltonians possessing magnetic anisotropy, which energetically favors certain directions in spin space.

Fullerene molecules are molecules of carbon that consist of a fixed number of 12 pentagons, and a number of hexagons which varies with the number of vertices $N$ as $\frac{N}{2}-10$ \cite{Fowler95}. The polygons share edges while each vertex is threefold coordinated. Due to the presence of pentagons the molecular topology is frustrated \cite{Ramirez94,Schnack10}, as even an isolated pentagon can not have antiferromagnetically interacting nearest-neighbor classical spins on its vertices pointing in antiparallel directions in its lowest energy configuration. The reason is the odd number of vertices and the closed boundary conditions. Since the pentagons are the source of frustration, it decreases on the average with $N$.

Perhaps the most representative fullerene molecule is the truncated icosahedron \cite{Kroto87,Smith98,Zhang07,Tamai04,Zeng01,Chen15,Manaa09}, which has the spatial symmetry of the icosahedral group $I_{h}$ \cite{Altmann94}. The truncated icosahedron becomes superconducting when doped with alkali metals \cite{Hebard91}. From a theoretical point of view, a first approximation has three of each carbon atom's valence electrons forming three $\sigma$-bonds using $sp^2$ hybrid orbitals, with the fourth valence electron delocalized.
In this way each carbon has one active radial p-orbital. When the nearest neighbor Hubbard hopping $t$ is much weaker than the long-range Coulomb or the on-site Hubbard repulsion $U$ \cite{Auerbach98,Fazekas99}, there is essentially one electron in each orbital and the nearest-neighbor interaction is antiferromagnetic \cite{Coffey92}. According to estimates of the on-site repulsion for the truncated icosahedron the molecule belongs to the intermediate $U$ regime of the Hubbard model \cite{Chakravarty91,Stollhoff91}. A first approximation is to consider the large $U$ limit of the Hubbard model, the AHM,
in order to calculate the influence of the frustrated connectivity on the magnetic properties.

For $I_h$-symmetry fullerene molecules the classical lowest energy configuration of the AHM has been found to have two magnetization gaps in an external field, the only exception being the smallest member of the class, the dodecahedron \cite{Prinzbach00,Wang01,Iqbal03}, where it has three \cite{Coffey92,NPK05,NPK07}. Such discontinuities are unexpected for a model lacking magnetic anisotropy, and originate in the frustrated connectivity of the molecules. In the opposite limit where the individual spin magnitude $s=\frac{1}{2}$ and 1, the dodecahedron has respectively one and two ground-state discontinuities in a field \cite{NPK05}. A high-field magnetization gap in the ground state was established to be a common feature of the $I_h$ fullerenes for $s=\frac{1}{2}$ \cite{NPK07}, while relatively small fullerene molecules of other symmetries have only pronounced magnetization plateaus in the full quantum limit \cite{NPK09}. It is noted here that further evidence for the validity of the AHM as a good description of the Hubbard model for the $I_h$ fullerenes was found in numerical calculations of the on-site repulsion $U$, which is stronger for the dodecahedron than the truncated icosahedron \cite{Lin07,Lin08}.

The classical ground-state magnetic response within the framework of the AHM has also been calculated for fullerene molecules resembling capped carbon nanotubes \cite{NPK17}. In this case the magnetic response depends on the nanotube chirality and the spatial symmetry. Armchair carbon nanotubes which are capped with non-neighboring pentagons and have $D_{5d}$ spatial symmetry possess a number of magnetization discontinuities which increases with their size. This occurs even though the only source of frustration are two groups of six pentagons located at the ends of the molecules, which furthermore become more strongly outnumbered as the clusters are filled in the middle with more unfrustrated hexagons with increasing size. For the cluster with 180 vertices there are already seven magnetization and one susceptibility discontinuities. Contrary to that, similar molecules with the slightly different $D_{5h}$ spatial symmetry reach a limit of one magnetization and two susceptibility lowest energy configuration discontinuities. Fullerene molecules resembling zigzag carbon nanotubes capped by neighboring pentagons also reach a fixed number of discontinuities with their size.


The above results demonstrate strong correlations between spatial symmetry and magnetic response for fullerene molecules. Such correlations have also been found between the dodecahedron and the icosahedron \cite{NPK05}, both of them being Platonic solids \cite{Plato}, with the latter the smallest cluster with $I_h$ symmetry which is however not a fullerene and is made up only of triangles \cite{Schroeder05,NPK15,Vaknin14,Engelhardt14,Hucht11,Sahoo12-1,Sahoo12,Strecka15,Karlova16,Karlova16-1,Karlova16-2}. It is noted here that the inclusion of higher-order exchange terms significantly enhances the discontinuous magnetic response \cite{NPK16-1,NPK17-1}.

Motivated by the rich magnetic properties of the different topology families of fullerene molecules, in this paper we calculate the ground-state magnetic response of molecules with $N \leq 36$. These are the most accessible molecules with respect to computational requirements. Discontinuities in the magnetization and the susceptibility are sought after, as well as correlations between cluster topology and magnetization response. This is first done for all the molecules with $N \leq 36$ for classical spins, which are mounted on the vertices of the molecules and interact according to the AHM. For some of the molecules the calculation is repeated for $s=\frac{1}{2}$. Due to the smallness of $N$ the 12 pentagons are never isolated, enhancing the effect of frustration.
It is found that the small fullerenes support classical ground-state magnetization and susceptibility discontinuities, providing solid evidence that the zero-temperature discontinuous magnetic response is a generic feature of fullerene molecules irrespective of spatial symmetry at the classical level of the AHM. Again, this is something not expected in the absence of magnetic anisotropy, with the origin of the discontinuities lying in the frustrated connectivity of the molecules. Two of them possess the most discontinuous classical response, with the first having $N=26$ and a magnetization curve with four magnetization and two susceptibility ground-state jumps. The second is an isomer of $N=34$ that has three jumps of each type. Comparing the number of discontinuities with the corresponding numbers for the $I_h$-symmetry and the nanotube-type molecules \cite{NPK05,NPK07,NPK17}, the $N \leq 36$ fullerenes can have the most discontinuous classical magnetic response, taking the cluster size into account. Another consequence of frustration is that for many molecules the classical ground state has finite magnetization in zero field, something unexpected for an even number of spins interacting antiferromagnetically and where each spin has the same number of nearest-neighbors. This is a purely classical effect, as a quantum system would have a degenerate ground state and no net magnetization. Small fullerenes come in different symmetries and topologies and there are only a few patterns of magnetic behavior that can be detected from such a relatively small sample. This is in contrast to the fullerenes families mentioned earlier which share the symmetry and topology. Larger clusters with the same symmetries with the $N \leq 36$ clusters can provide more insight on correlations between magnetic response and topology. The paper provides more evidence that fullerene molecules generically support discontinuous ground-state magnetic response, as the classical discontinuities are expected to survive at least for high $s$. Such magnetic behavior shows the possibility of fabricating small entities of the molecular nanomagnet type that can be tuned between well-separated magnetization values by weak variations of an external field.

The $s=\frac{1}{2}$ ground-state magnetic response was calculated for five of the $N \leq 36$ molecules. No magnetization jumps were found but only magnetization plateaux, which is in agreement with the available results for six other small fullerene molecules \cite{NPK09}, showing that the quantum fluctuations work against the classical discontinuities. Similarly to the classical case, there is no pattern of magnetic behavior that can be detected. It is concluded that the fullerenes with icosahedral symmetry are the only ones known supporting ground-state magnetization discontinuities for $s=\frac{1}{2}$. It is also found that the 6th isomer of $N=34$ has a doubly-degenerate ground state. This was also the case for the $T_d$ isomer of $N=28$. These two clusters have the biggest zero-field magnetization in the classical case. They point to the fact that small structural distortions can reduce the symmetry and weakly split the ground-state doublet, something reminiscent of single molecule magnets \cite{Gatteschi06}, which have been proposed as potential qubits in quantum computers \cite{Leuenberger01}.

The plan of this paper is as follows: Section \ref{sec:model} introduces the model, while Sec. \ref{sec:classicalspins} describes the classical lowest energy configuration in zero field and the classical magnetization response in an external field. Section \ref{sec:spinsonehalf} presents the results for $s=\frac{1}{2}$, and Sec. \ref{sec:conclusions} the conclusions.

\section{Model}
\label{sec:model}
The fullerene molecules with $N \leq 36$ are shown in Ref. \cite{Fowler95}. They are distinguished by $N$ and by an index characterizing the different isomers. The Hamiltonian of the AHM for spins $\vec{s}_i$ and $\vec{s}_j$ mounted on the vertices $i,j=1,\dots,N$ of the fullerene molecules is
\begin{eqnarray}
H & = & J \sum_{<ij>} \textrm{} \vec{s}_i \cdot \vec{s}_j - h \sum_{i=1}^{N} s_i^z
\label{eqn:Hamiltonian}
\end{eqnarray}
$J$ is the strength of the exchange interaction, which is set to 1, defining the unit of energy. $<ij>$ indicates that interactions are limited to nearest neighbors. The magnetic field has strength $h$ and is taken to be directed along the $z$ axis. In Hamiltonian (\ref{eqn:Hamiltonian}) the minimization of the exchange energy competes with the one of the magnetic energy, with the frustrated topology of the molecules playing a very important role.

At the classical level, numerical minimization of the Hamiltonian gives the lowest energy and the corresponding spin configuration as a function of $h$ \cite{Coffey92,NPK05,NPK07,NPK13,NPK15,NPK15-1,NPK16,NPK16-1,NPK17,NPK17-1,Machens13}. Each spin $\vec{s}_i$ is a classical unit vector defined by a polar $\theta_i$ and an azimuthal $\phi_i$ angle. A random initial configuration of the spins is chosen for a specific $h$ and each angle is moved opposite its gradient direction, until the minimum of the energy is reached. Repetition of the procedure for different initial configurations ensures that the absolute lowest energy configuration is found for every $h$.

In the quantum-mechanical case the Hamiltonian is diagonalized according to its spatial and spin symmetries \cite{NPK04,NPK05,NPK07,NPK09,NPK16,NPK15,NPK15therm,NPK16therm,Machens13}. Then a characterization of the eigenstates according to their symmetry properties is possible, together with the minimization of memory requirements for diagonalization.
Hamiltonian (\ref{eqn:Hamiltonian}) commutes with the total spin $S$ and its projection along the $z$ axis $S^z$,
and the method for the symmetry characterization of the eigenstates is most easily carried out within each individual $S^z$ sector. Hamiltonian (\ref{eqn:Hamiltonian}) is symmetric under combinations of spin permutations that respect the connectivity of a cluster. The group of permutations is the symmetry group of the cluster in real space \cite{Altmann94}. Hamiltonian (\ref{eqn:Hamiltonian}) also posseses time-reversal symmetry, and invertion of the spins is a symmetry operation in the $S^z=0$ sector. The corresponding group is comprised of the identity and the spin inversion operation. The full symmetry group of the Hamiltonian is the product of the real space and the spin inversion group. Taking the full symmetry into account, the $S^z$-basis states can be projected into states that transform under specific irreducible representations of the full symmetry group. In this way the Hamiltonian is block diagonalized into smaller matrices characterized by symmetry, and simultaneously the maximal matrix dimension is dramatically reduced compared to the full $S^z$-subspace size. Then Lanczos diagonalization in all the irreducible representations produces by comparison the lowest energy in every $S^z$ sector, and the magnetization response in a field is calculated.
Comparison of the lowest energies in the different $S^z$ sectors enables the characterization of the corresponding eigenstates by $S$. It is noted that degeneracies are reported with respect to multiplets of eigenstates with a specific value of $S$, and that each of these multiplets corresponds to a number of $2S+1$ eigenstates, each having a different $S^z$ value.

\section{Classical Spins}
\label{sec:classicalspins}

\subsection{Zero Magnetic Field}
\label{subsec:zeromagneticfield}

\begin{figure}
\includegraphics[width=3.4in,height=2.7in]{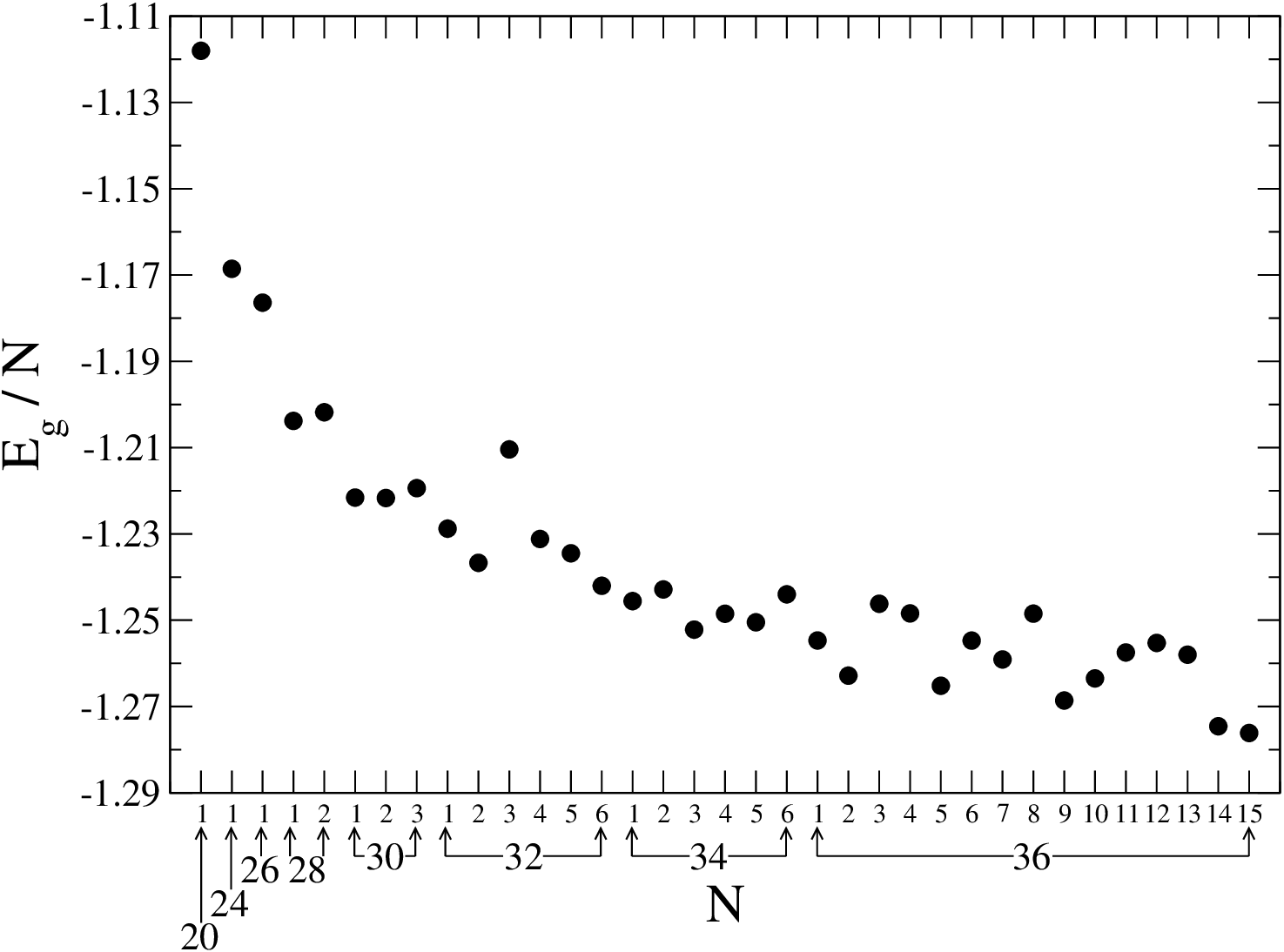}
\caption{Classical ground-state energy per spin $\frac{E_g}{N}$ as a function of $N$ (Table \ref{table:1}). The different isomers are specifically numbered for each $N$.
}
\label{fig:zerofieldenergy}
\end{figure}

Table \ref{table:1} lists the zero-field ground-state energy per spin $\frac{E_g}{N}$ of Hamiltonian (\ref{eqn:Hamiltonian}). It decreases on the average with $N$ as also seen in Fig. \ref{fig:zerofieldenergy}, as the number of hexagons increases and frustration on the average decreases. $\frac{E_g}{N}$ approaches more closely with $N$ the value $-\frac{3}{2}$, the one of the hexagonal lattice.


Table \ref{table:1} also lists the zero-field ground-state magnetization per spin $\frac{M_g}{N}$. This is generally expected to be zero for antiferromagnetic interactions in the absence of a field, especially since fullerene molecules have even $N$ and each spin has the same number of nearest-neighbors. On the contrary, many molecules have finite zero-field magnetization, showing again the importance of the frustrated fullerene topology and its lack of bipartiteness. A residual magnetization relates to a magnetized entity even in the absence of a field. As the dodecahedron has been shown to have enriched magnetic response in the presence of intermolecular interactions without having zero-field magnetization \cite{NPK16}, it would be of interest to examine the behavior of interacting collections of fullerene molecules with $\frac{M_g}{N} \neq 0$. Their magnetization could demonstrate even-odd effects, as is the case with an open spin chain \cite{Machens13,NPK15-1}, but here in the number of molecules and on top of the isolated molecule properties.

The clusters with finite zero-field magnetization possess in general the lowest symmetry among their isomers, while for $N=34$ all isomers have residual magnetization. The $N=28$ cluster with $T_d$ symmetry (Fig. 1(c) in Ref. \cite{NPK09}) has the largest value of $\frac{M_g}{N}$. In Ref. \cite{NPK09} it was shown that it has a doubly-degenerate ground state for $s=\frac{1}{2}$ as well as quite a small singlet-triplet gap, indicating that the large residual magnetization leaves its fingerprint down to low $s$. Along the same lines in the same reference it was shown that the singlet-triplet gap of the $N=26$ molecule (Fig. 1(b) in Ref. \cite{NPK09}) is quite bigger, while here its classical residual magnetization is found to be roughly an order of magnitude smaller than the one of the $N=28$ molecule with $T_d$ symmetry. The correlation between large classical residual magnetization and small singlet-triplet gap is further supported by the results of Sec. \ref{sec:spinsonehalf}, by the $N=34$ cluster with $C_{3v}$ symmetry that has the second largest zero-field magnetization and the 3rd isomer with $N=30$ which also has a significant finite $\frac{M_g}{N}$.

\begin{figure}
\includegraphics[width=3.4in,height=2.7in]{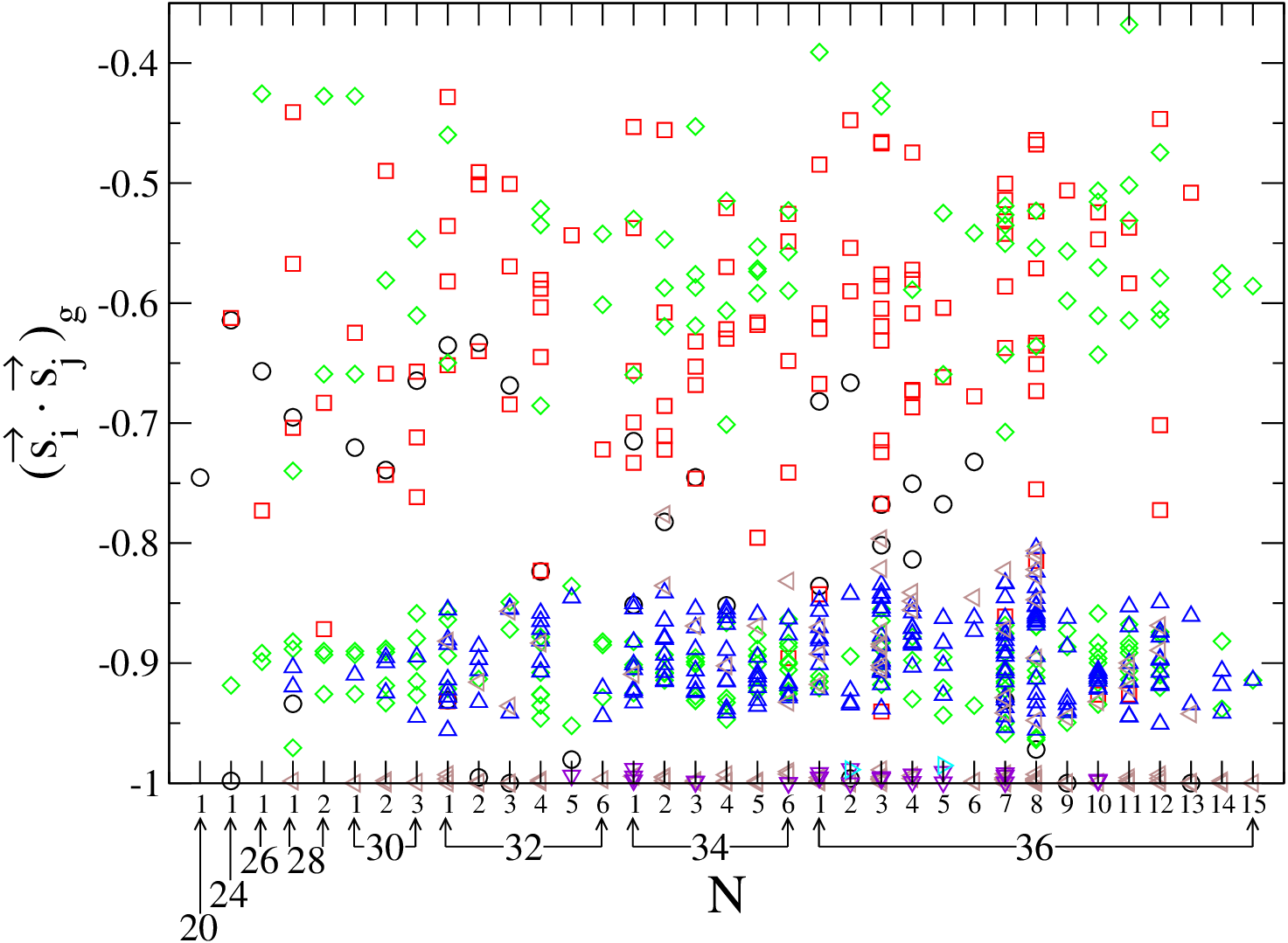}
\caption{(Color online) Unique nearest-neighbor correlations $(\vec{s}_i \cdot \vec{s}_j)_g$ in the zero-field lowest energy state for classical spins. The different isomers are specifically numbered for each $N$. The (black) circles show correlations between spins belonging only to pentagons, the (red) squares between a spin belonging only to pentagons and one belonging to two pentagons, the (green) diamonds between spins belonging to two pentagons, the (blue) up triangles between a spin belonging to two and a spin belonging to one pentagon, the (brown) left triangles between spins belonging to one pentagon, the (violet) down triangles between a spin belonging to one pentagon and a spin that belongs only to hexagons, and the (cyan) right triangles between spins that belong only to hexagons.
}
\label{fig:zerofieldnncorr}
\end{figure}

The vertices of the fullerene molecules are threefold coordinated and participate in three different polygons, which can be pentagons and hexagons. The only molecule where all vertices and consequently the spins mounted on them are equivalent is the dodecahedron ($N=20$), which consists only of pentagons. In general each spin can belong to a number of pentagons ranging from three down to zero. Fig. \ref{fig:zerofieldnncorr} shows the unique nearest-neighbor correlations in the zero-field ground state $(\vec{s}_i \cdot \vec{s}_j)_g$. For the $N=20$ molecule there is only one unique nearest-neighbor correlation due to the equivalence of its sites. In general the correlations become more antiferromagnetic the less pentagons and more hexagons the spins belong to. The reason is that an isolated pentagon is frustrated, in contrast to the unfrustrated isolated hexagon. There are only some exceptions where nearest-neighbors solely belonging to pentagons develop very strong antiferromagnetic correlations.

\subsection{Magnetization Response}
\label{subsec:magnetizationresponse}

Table \ref{table:1} lists the saturation field $h_{sat}$ of the $N \leq 36$ fullerene molecules, which increases on the average with $N$. This is because the number of hexagons increases with $N$ and since they are unfrustrated they support stronger antiferromagnetic nearest-neighbor correlations, as already seen in Sec. \ref{subsec:zeromagneticfield}, which necessitate a stronger field to align all the spins along its direction.

\begin{figure}
\includegraphics[width=3.4in,height=2.7in]{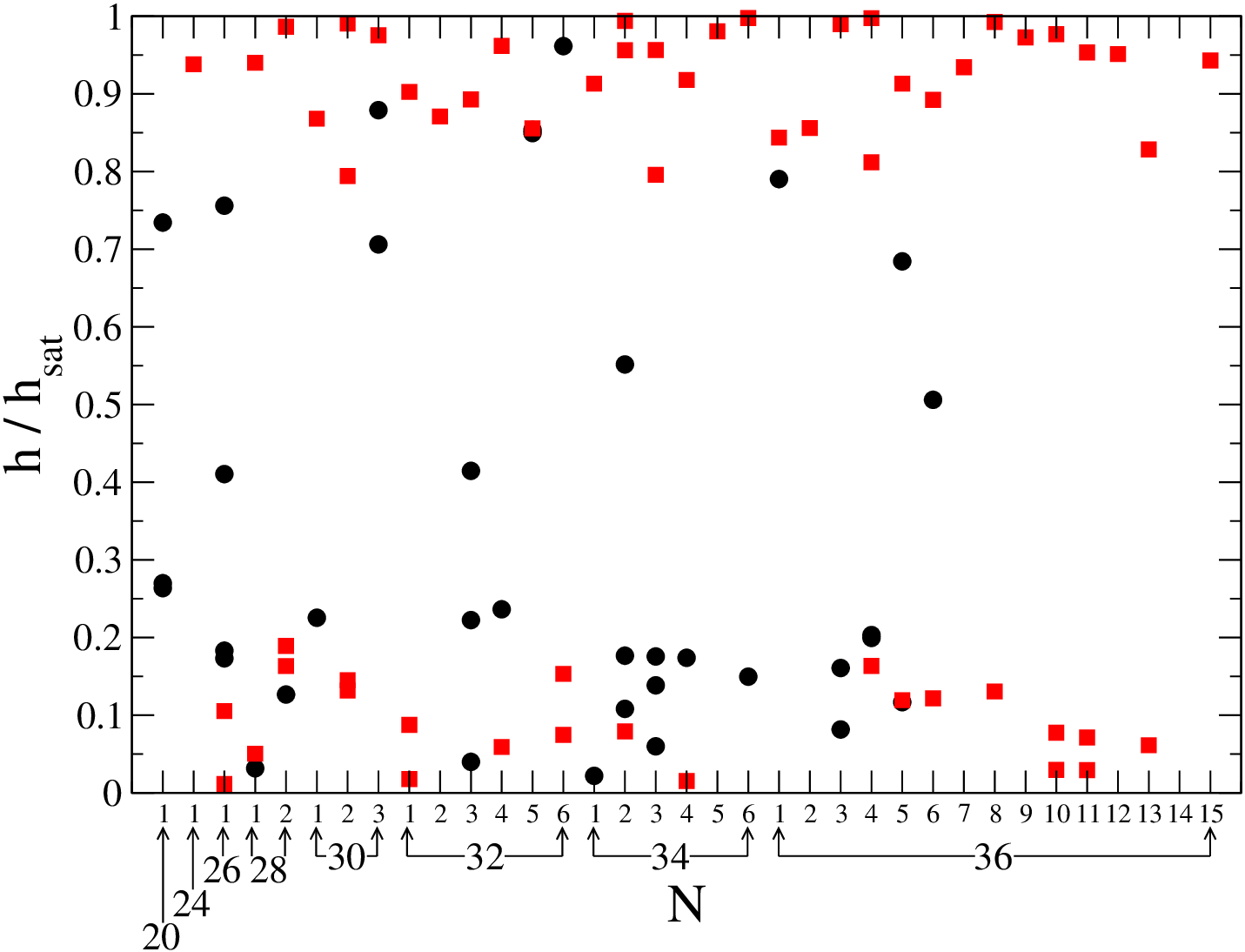}
\caption{(Color online) Discontinuity magnetic field values over the saturation field $h_{sat}$ for classical spins. The different isomers are specifically numbered for each $N$. The (black) circles correspond to the magnetization discontinuities (Table \ref{table:2}), and the (red) squares to the susceptibility discontinuities (Table \ref{table:3}).
}
\label{fig:disc}
\end{figure}

\begin{figure}
\includegraphics[width=3.4in,height=2.7in]{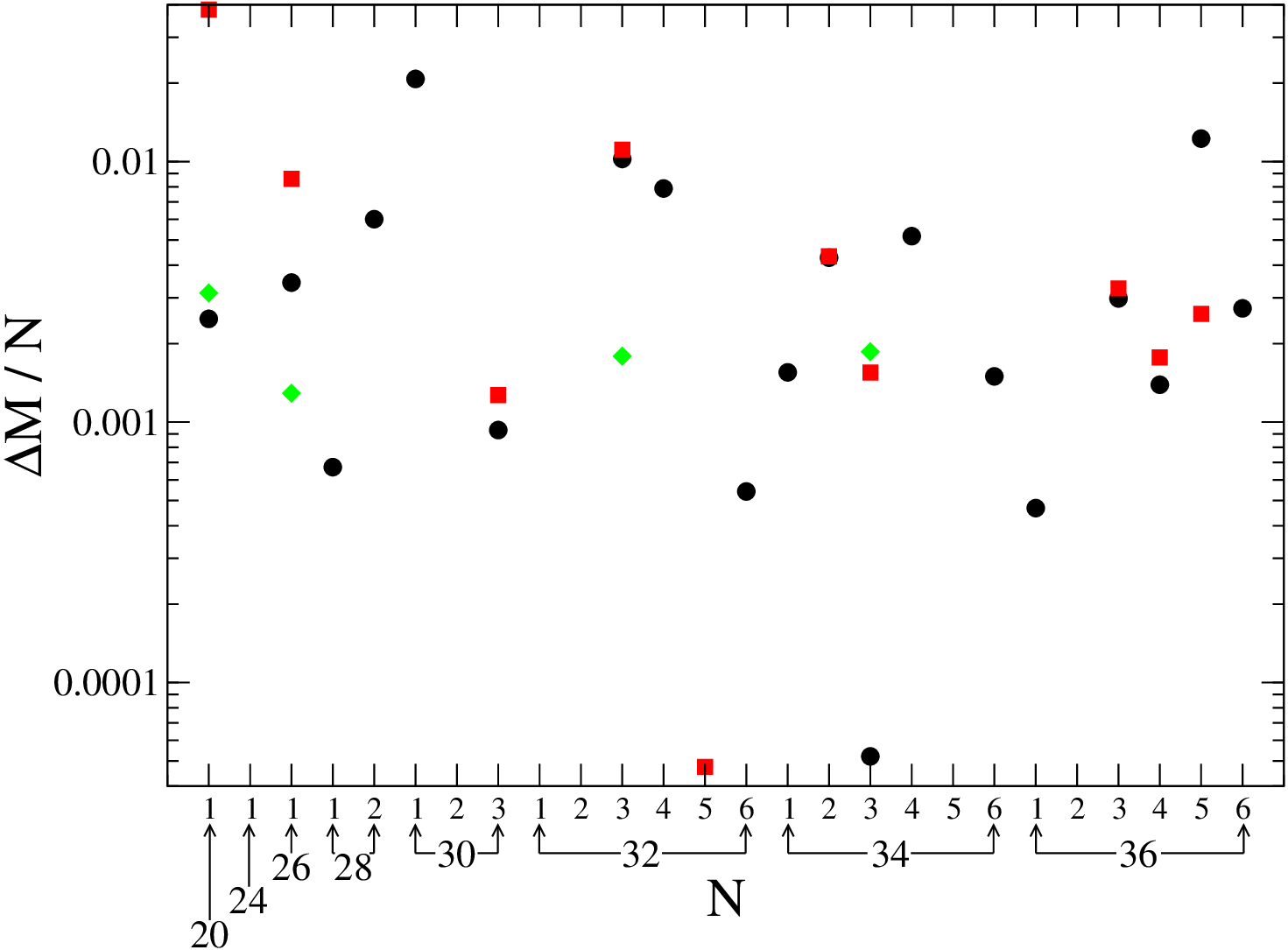}
\caption{(Color online) Magnetization discontinuity width per spin $\frac{\Delta M}{N}$ for the fields shown in Fig. \ref{fig:disc} (Table \ref{table:2}) for classical spins. The different isomers are specifically numbered for each $N$. The (black) circles correspond to the first magnetization discontinuity for each molecule, the (red) squares to the second, and the (green) diamonds to the third.
}
\label{fig:magndiscwidth}
\end{figure}

Fig. \ref{fig:disc} plots the magnetization and susceptibility ground-state discontinuities in an external field, listed in Tables \ref{table:2} and \ref{table:3}. The widths of the magnetization gaps are plotted in Fig. \ref{fig:magndiscwidth}. The results for the dodecahedron ($N=20$) have already been presented \cite{NPK05,NPK07}. A common pattern is that most of the discontinuities occur for fields small or close to saturation, which has been explained by the stronger sensitivity of the pentagon spins to the field, originating in the frustrated nature of the pentagons \cite{NPK17}. This results in weaker zero-field antiferromagnetic correlations within the pentagons in comparison with the hexagons, as already seen in Sec. \ref{subsec:zeromagneticfield}.
The largest number of discontinuities have the $N=26$ molecule, four of the magnetization and two of the susceptibility, and the second $N=34$ isomer, which has three each.

It is not obvious how to determine the number of discontinuities for a specific cluster without doing the full numerical calculation. Patterns of magnetic behavior have been detected for fullerene molecules belonging to specific spatial symmetry and connectivity subsets, like molecules with $I_h$ symmetry or armchair and zigzag nanotube-type molecules capped at their ends with pentagons \cite{NPK05,NPK07,NPK17}. Here $N$ is relatively small and such an insight could be possible when the calculation is extended to bigger molecules that are structural relatives of the $N \leq 36$ clusters. Still, since the frustrated pentagons outnumber the hexagons when $N$ is small their influence is stronger in comparison with the case of bigger molecules and more discontinuous magnetic response is expected. Common magnetization patterns can still be detected in some cases.

A common pattern of magnetization in an external field has been demonstrated for the family of fullerene molecules that have the shape of (5,0) zigzag carbon nanotubes and are capped at their ends with six neighboring pentagons, which form one half of a dodecahedron \cite{NPK17}. These molecules have a number of vertices which is a multiple of 10, having $D_{5d}$ symmetry when it is an even and $D_{5h}$ symmetry when it is an odd multiple of 10 \cite{Fowler95}. The first isomer with $N=30$ (Fig. 1(d) in Ref. \cite{NPK09}) is the smallest member of this family with $D_{5h}$ symmetry, having also a low-field magnetization and a high-field susceptibility discontinuity whose corresponding magnetic fields fit with the ones of the rest of the family (Tables \ref{table:2} and \ref{table:3}, and Fig. 7 in Ref. \cite{NPK17}). The body of the nanotube is minimal as it is formed by a single row of hexagons. The dodecahedron is the smallest member of this family, and has the special property that the complete lack of hexagons forces the two caps to share some of their spins. This results in a higher symmetry, described by the $I_h$ group, and three magnetization discontinuities (Table \ref{table:2} and Refs. \cite{NPK05,NPK07}). This shows according to the previous paragraph that when the pentagons are brought in closer contact with each other a more discontinuous response emerges with respect to the bigger clusters of the family.

Another family of fullerene molecules resembling carbon nanotubes has the form of a (6,0) zigzag nanotube capped at both ends by a closed ring of six adjacent pentagons forming a hexagon in the middle \cite{Fowler95}. $N$ is a multiple of 12 and there is a 6-fold rotational symmetry axis, with the symmetry being $D_{6d}$ when $N$ is an even and $D_{6h}$ when $N$ is an odd multiple of 12. The smallest two molecules of this family are the $N=24$ cluster (Fig. 1(a) in Ref. \cite{NPK09}) and the 15th isomer with $N=36$, which only have a high-field susceptibility discontinuity (Table \ref{table:3}). Similarly to the first isomer of $N=30$, for the $N=36$ isomer the nanotube body is the smallest possible, formed by a single row of hexagons. The $N=24$ cluster is similar to the dodecahedron, being the limit where there is no nanotube body and the two caps share some of their spins.

Finally isomer 1 of $N=32$ and isomers 10 and 11 of $N=36$ have the same magnetization response pattern, with two low-field and one high-field susceptibility discontinuities. The symmetry of these clusters is $C_2$.






\section{Quantum Spins $s=\frac{1}{2}$}
\label{sec:spinsonehalf}

The ground-state energy per spin when $s=\frac{1}{2}$ is listed in Table \ref{table:spinonehalf} for five of the $N \leq 36$ fullerene molecules. As in the classical case it decreases on the average with $N$ with frustration getting weaker. This was also the case for six other small fullerene molecules \cite{NPK09}.

\begin{figure}
\centerline{
\includegraphics[width=0.24\textwidth,height=1.4in]{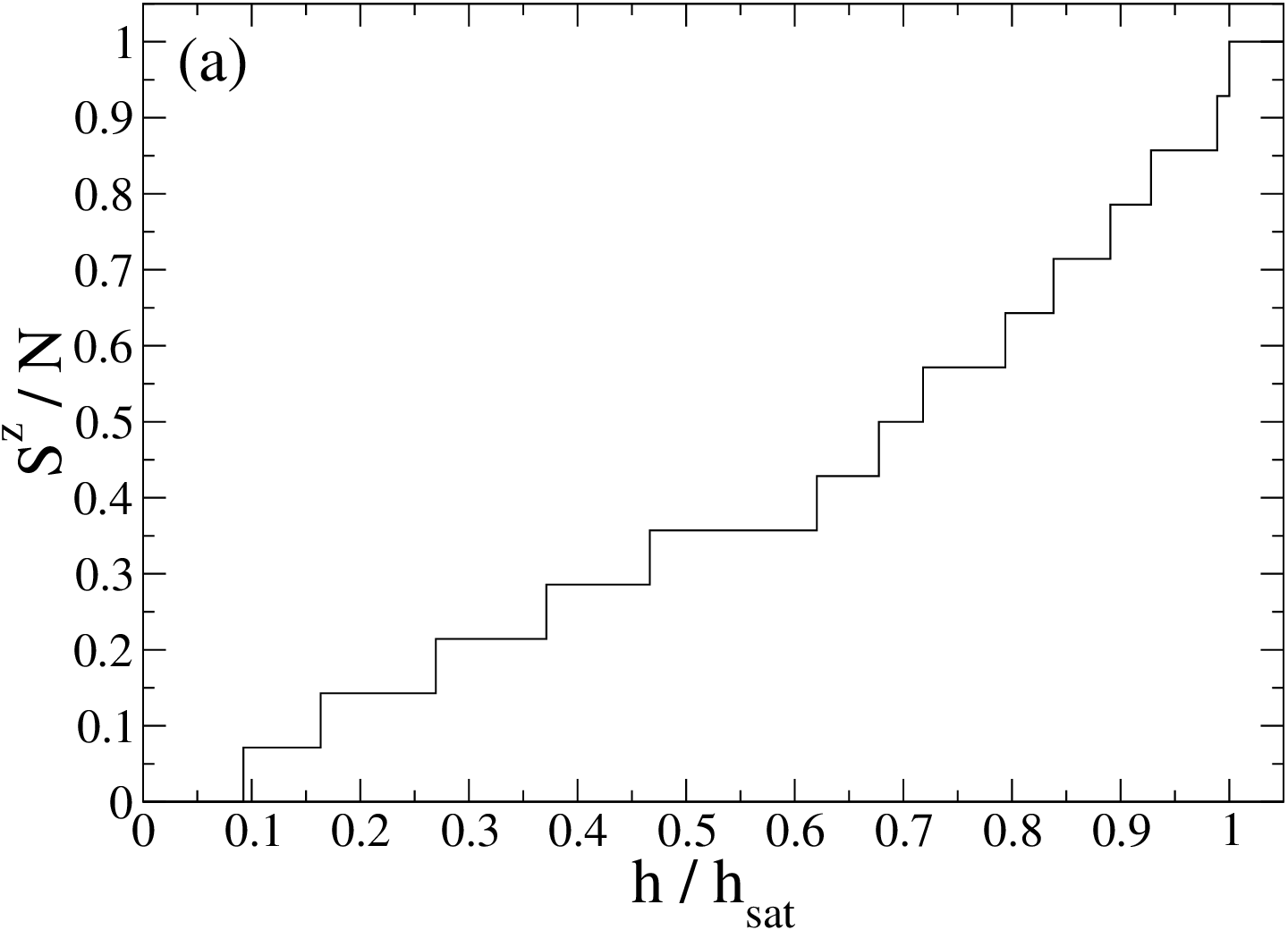}
\includegraphics[width=0.24\textwidth,height=1.4in]{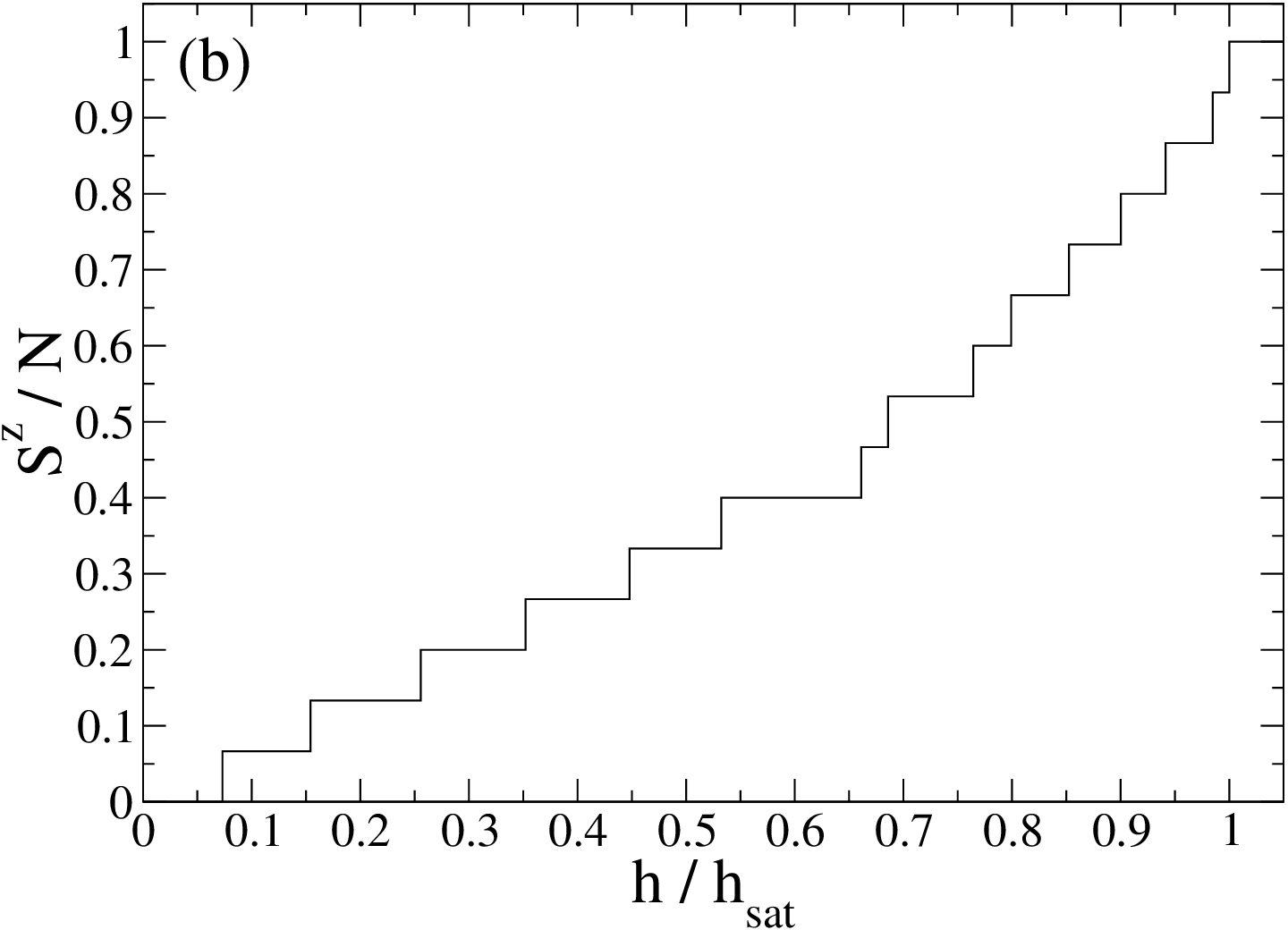}
}
\vspace{17pt}
\centerline{
\includegraphics[width=0.24\textwidth,height=1.4in]{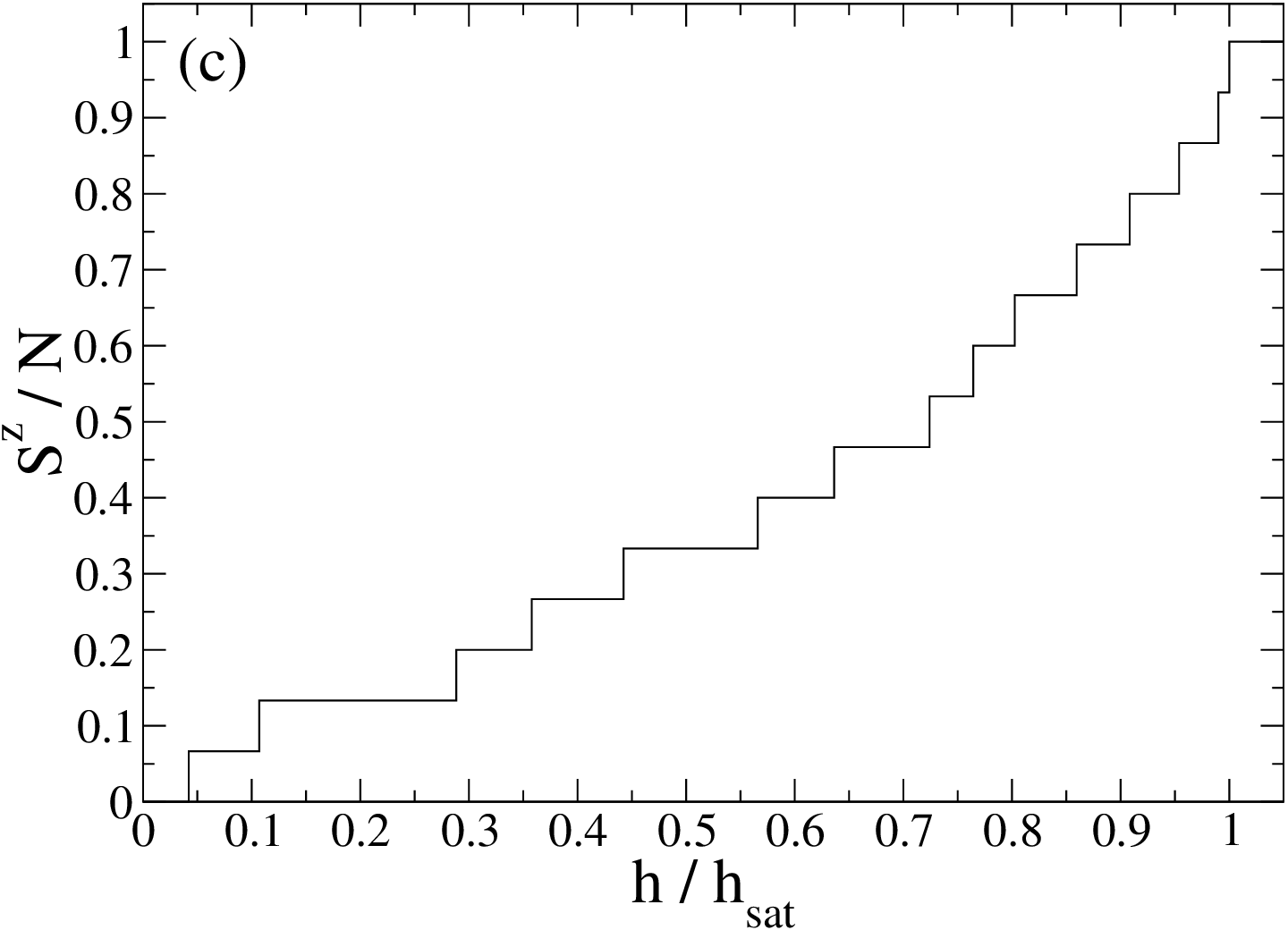}
\hspace{0pt}
\includegraphics[width=0.24\textwidth,height=1.4in]{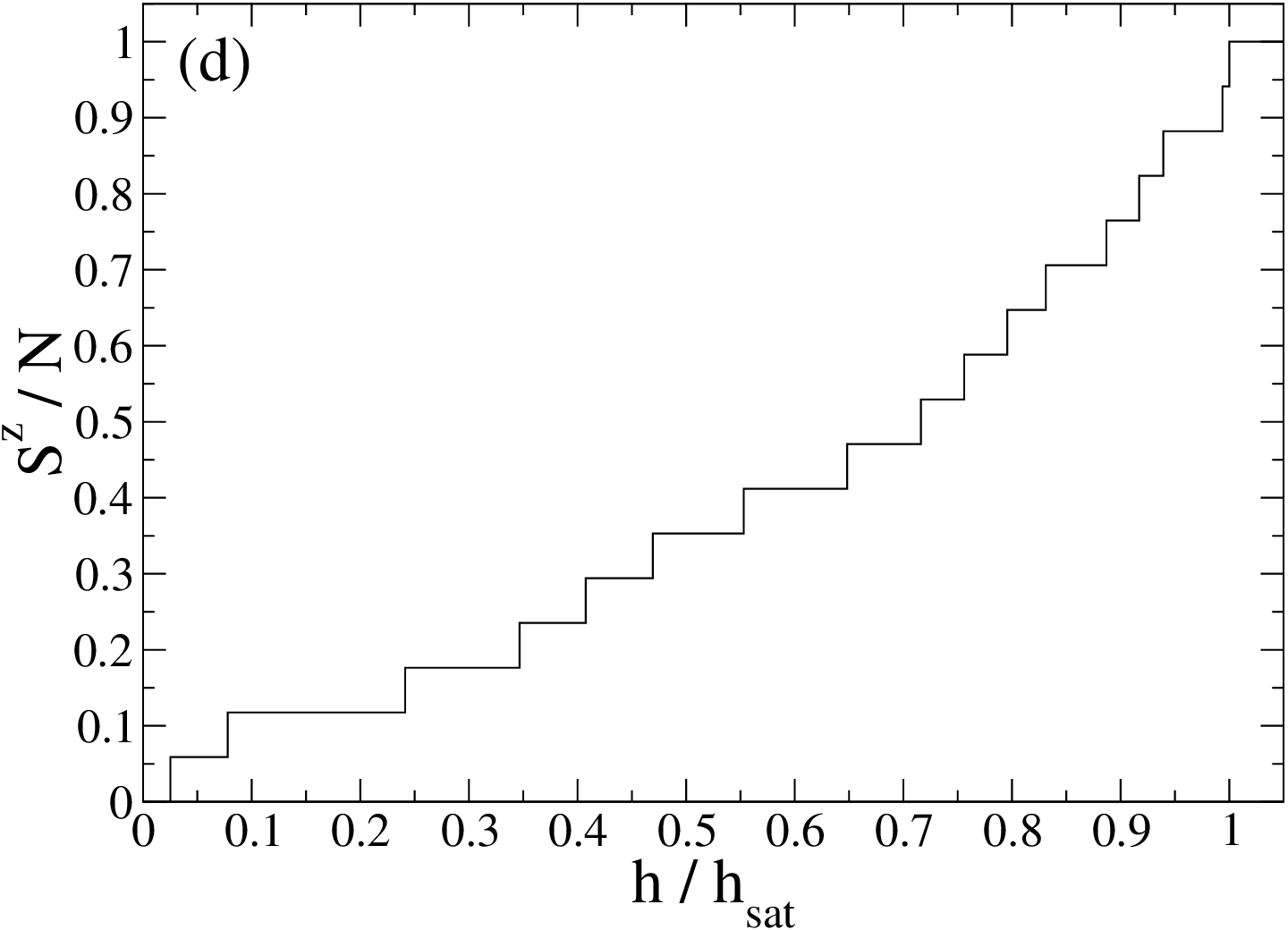}
}
\vspace{17pt}
\centerline{
\includegraphics[width=0.24\textwidth,height=1.4in]{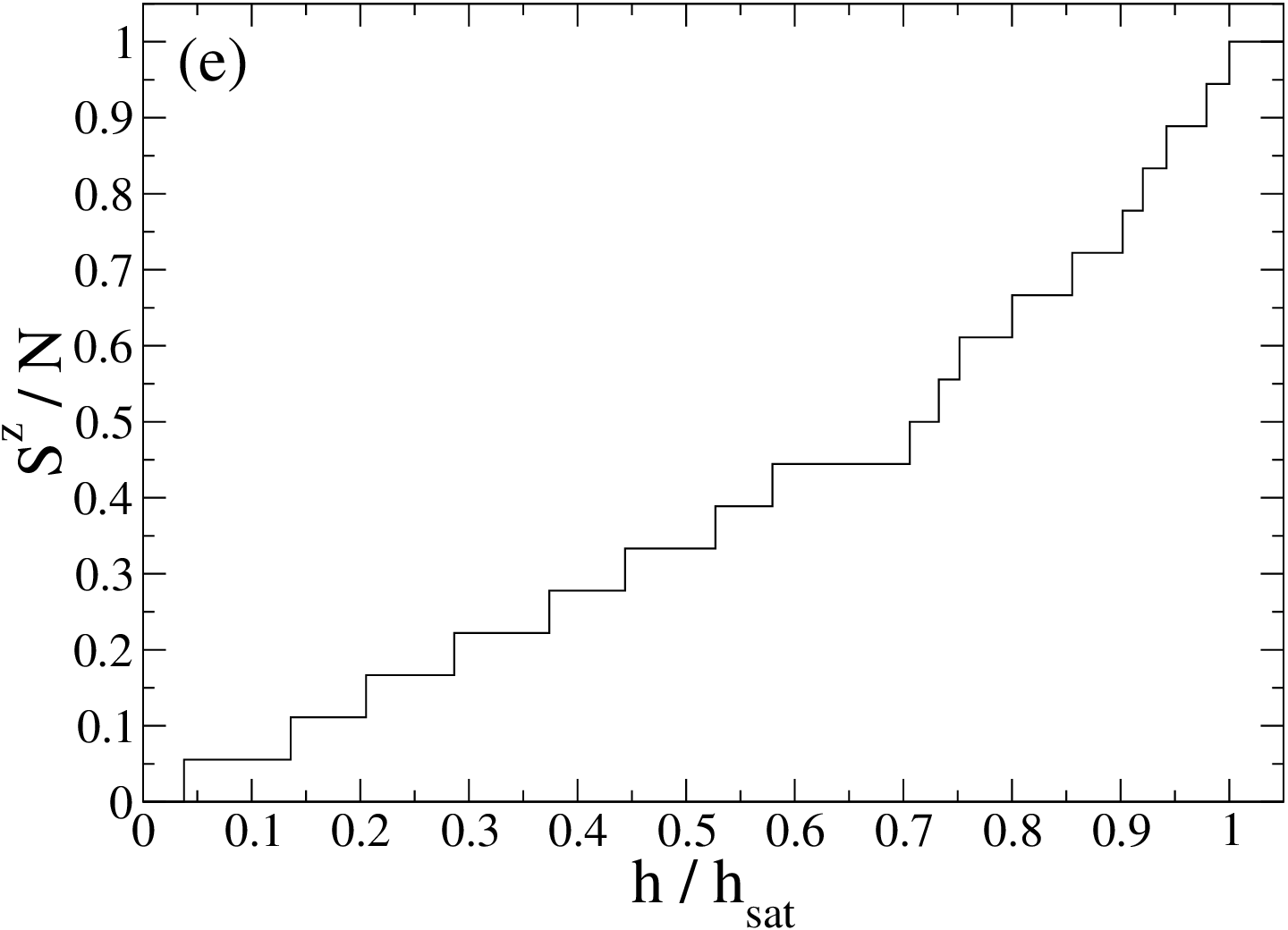}
}
\caption{Ground-state magnetization per spin $\frac{S^z}{N}$ as a function of the magnetic field $h$ over its saturation value $h_{sat}$ for isomer (Table \ref{table:spinonehalf}) (a) 1 with $N=28$, (b) 2 with $N=30$, (c) 3 with $N=30$, (d) 6 with $N=34$, and (e) 15 with $N=36$.
}
\label{fig:icosahedronconfigurationsfivedifferent}
\end{figure}

The lowest lying level of Hamiltonian (\ref{eqn:Hamiltonian}) in each $S$ sector along with its symmetry properties is listed in Table \ref{table:spinonehalfenergies} for the five fullerene molecules, while their ground-state magnetization response is plotted in Fig. \ref{fig:icosahedronconfigurationsfivedifferent}. There are no magnetization discontinuities but only magnetization plateaux, unlike the classical case. This shows that quantum fluctuations work against jumps, with the $I_h$ fullerene molecules the only ones known supporting magnetization gaps at the full quantum limit \cite{NPK05,NPK07}. Similar results have been found for the other six small fullerene molecules \cite{NPK09}.

The $N=34$ cluster has a doubly-degenerate ground state, as was also the case for the $T_d$ isomer of $N=28$ \cite{NPK09}. Even though the ground-state degeneracy can be found analytically only for the very limited number of Lieb-Mattis systems, from the point of view of experience a doubly-degenerate ground state is still something unexpected. These two clusters have the largest residual classical ground-state magnetizations (Table \ref{table:1}). Two other clusters with residual magnetizations which however have a non-degenerate ground state when $s=\frac{1}{2}$ are the $N=26$ cluster \cite{NPK09} and the third isomer with $N=30$. It is of interest to examine if other fullerene molecules with a finite zero-field classical ground-state magnetization have a degenerate quantum-mechanical ground state. In such a case it could be possible with a weak spatial symmetry breaking generated by a structural distortion to weakly split the ground-state doublet even in the absence of magnetic anisotropy. In the case of single molecule magnets a strong magnetic anisotropy along with a tunneling term generates an almost degenerate low-lying doublet \cite{Gatteschi06}, and single molecule magnets have been proposed as potential qubits in quantum computers \cite{Leuenberger01}.




\section{Conclusions}
\label{sec:conclusions}

The AHM has been considered for fullerene molecules with $N \leq 36$, which belong to various symmetry groups \cite{Fowler95}. At the classical level the ground-state magnetic response has many magnetization and susceptibility discontinuities, as many as six in total for a single molecule. This demonstrates that the discontinuities are a generic feature of the fullerene molecules irrespective of their size and symmetry, originating in the frustrated pentagons. Frustration also supports a finite zero-field ground-state magnetization for many molecules, something unexpected for antiferromagnetic interactions between an even number of spins which furthermore have the same number of nearest-neighbors. For $s=\frac{1}{2}$ only magnetization plateaux and no discontinuities are found for five of the molecules. This leaves the icosahedral symmetry fullerenes as the only ones known supporting magnetization discontinuities at the full quantum limit \cite{NPK05,NPK07}. One of the molecules has a doubly-degenerate ground state. Few symmetry patterns in the classical magnetic response are found for these relatively small and highly frustrated molecules. An extension of the investigation to bigger molecules is needed to fully determine the fullerene ground-state magnetization response, as well as to clearly identify its symmetry patterns \cite{NPK05,NPK07,NPK17}.

\bibliography{smallfullerenes}

\begin{table}[h]
\begin{center}
\caption{Number of vertices $N$, isomer index no., spatial symmetry group, zero-field ground-state energy per spin $\frac{E_g}{N}$, zero-field ground-state residual magnetization per spin $\frac{M_g}{N}$, and saturation magnetic field $h_{sat}$ for classical spins.}
\begin{tabular}{c|c|c|c|c|c|c|c|c|c|c|c}
$N$ & no. & group & $\frac{E_g}{N}$ & $\frac{M_g}{N}$ & $h_{sat}$ & $N$ & no. & group & $\frac{E_g}{N}$ & $\frac{M_g}{N}$ & $h_{sat}$ \\
\hline
20 & 1 & $I_{h}$ & -1.11803 & 0 & 3 + $\sqrt{5}$ & 34 & 5 & $C_2$ & -1.25048 & $4.53102 \times 10^{-3}$ & 5.57316 \\
\hline
24 & 1 & $D_{6d}$ & -1.16856 & 0 & 4 + $\sqrt{2}$ & 34 & 6 & $C_{3v}$ & -1.24399 & $1.22732 \times 10^{-2}$ & 5.58911 \\
\hline
26 & 1 & $D_{3h}$ & -1.17639 & $3.67801 \times 10^{-3}$ & 4 + $\sqrt{2}$ & 36 & 1 & $C_2$ & -1.25469 & $2.55969 \times 10^{-3}$ & 5.67022 \\
\hline
28 & 1 & $D_2$ & -1.20381 & 0 & 5.52469 & 36 & 2 & $D_2$ & -1.26285 & 0 & 5.70573 \\
\hline
28 & 2 & $T_{d}$ & -1.20181 & $1.58661 \times 10^{-2}$ & 4 + $\sqrt{2}$ & 36 & 3 & $C_1$ & -1.24618 & $6.98103 \times 10^{-3}$ & 5.62702 \\
\hline
30 & 1 & $D_{5h}$ & -1.22157 & 0 & 3 + $\sqrt{7}$ & 36 & 4 & $C_s$ & -1.24841 & $3.51335 \times 10^{-3}$ & 5.62854 \\
\hline
30 & 2 & $C_{2v}$ & -1.22168 & 0 & 5.55893 & 36 & 5 & $D_2$ & -1.26518 & 0 & 5.65106 \\
\hline
30 & 3 & $C_{2v}$ & -1.21938 & $8.40221 \times 10^{-3}$ & 5.52183 & 36 & 6 & $D_{2d}$ & -1.25471 & 0 & 5.61106 \\
\hline
32 & 1 & $C_2$ & -1.22877 & $3.81513 \times 10^{-3}$ & 5.59819 & 36 & 7 & $C_1$ & -1.25909 & $5.22764 \times 10^{-3}$ & 5.63238 \\
\hline
32 & 2 & $D_2$ & -1.23670 & 0 & 5.62965 & 36 & 8 & $C_s$ & -1.24846 & $5.38740 \times 10^{-3}$ & 5.61100 \\
\hline
32 & 3 & $D_{3d}$ & -1.21041 & 0 & $\frac{9+\sqrt{5}}{2}$ & 36 & 9 & $C_{2v}$ & -1.26860 & 0 & 5.63819 \\
\hline
32 & 4 & $C_2$ & -1.23116 & $6.91413 \times 10^{-3}$ & 5.55841 & 36 & 10 & $C_2$ & -1.26348 & $9.87385 \times 10^{-3}$ & 5.61435 \\
\hline
32 & 5 & $D_{3h}$ & -1.23449 & 0 & 5.61050 & 36 & 11 & $C_2$ & -1.25747 & $8.24750 \times 10^{-3}$ & 5.60673 \\
\hline
32 & 6 & $D_3$ & -1.24202 & 0 & 5.51345 & 36 & 12 & $C_2$ & -1.25527 & $4.95511 \times 10^{-3}$ & 5.61051 \\
\hline
34 & 1 & $C_2$ & -1.24557 & $4.40400 \times 10^{-3}$ & 5.63794 & 36 & 13 & $D_{3h}$ & -1.25800 & 0 & 5.67513 \\
\hline
34 & 2 & $C_s$ & -1.24286 & $6.27599 \times 10^{-3}$ & 5.58894 & 36 & 14 & $D_{2d}$ & -1.27458 & 0 & 5.58504 \\
\hline
34 & 3 & $C_s$ & -1.25218 & $4.42197 \times 10^{-3}$ & 5.59535 & 36 & 15 & $D_{6h}$ & -1.27614 & 0 & 3 + $\sqrt{7}$ \\
\hline
34 & 4 & $C_2$ & -1.24850 & $1.80822 \times 10^{-3}$ & 5.61420 & & & & &
\end{tabular}
\label{table:1}
\end{center}
\end{table}

\begin{table}[h]
\begin{center}
\caption{Number of vertices $N$, isomer index no., spatial symmetry group, magnetic field $h_M$ of the magnetization discontinuity with respect to the saturation field $h_{sat}$, and magnetization per spin below ($M_-/N$) and above ($M_+/N$) the discontinuity for classical spins.}
\begin{tabular}{c|c|c|c|c|c|c|c|c|c|c|c}
$N$ & no. & group & $h_M / h_{sat}$ & $M_- / N$ & $M_+ / N$ & $N$ & no. & group & $h_M / h_{sat}$ & $M_- / N$ & $M_+ / N$ \\
\hline
20 & 1 & $I_{h}$ & 0.26350 & 0.22411 & 0.22660 &  32 & 6 & $D_3$ & 0.96137 & 0.96430 & 0.96485 \\
\hline
20 & 1 & $I_{h}$ & 0.26983 & 0.23688 & 0.27518 & 34 & 1 & $C_2$ & 0.021911 & 0.021984 & 0.023535 \\
\hline
20 & 1 & $I_{h}$ & 0.73428 & 0.74766 & 0.75079 & 34 & 2 & $C_s$ & 0.10814 & 0.096625 & 0.10090 \\
\hline
26 & 1 & $D_{3h}$ & 0.17314 & 0.15219 & 0.15562 & 34 & 2 & $C_s$ & 0.17662 & 0.17472 & 0.17905 \\
\hline
26 & 1 & $D_{3h}$ & 0.18293 & 0.17088 & 0.17947 & 34 & 2 & $C_s$ & 0.55157 & 0.56518 & 0.56521 \\
\hline
26 & 1 & $D_{3h}$ & 0.41039 & 0.41809 & 0.41938 & 34 & 3 & $C_s$ & 0.059914 & 0.050270 & 0.050322 \\
\hline
26 & 1 & $D_{3h}$ & 0.75599 & 0.77098 & 0.77207 & 34 & 3 & $C_s$ & 0.13836 & 0.12594 & 0.12749 \\
\hline
28 & 1 & $D_2$ & 0.031510 & 0.031128 & 0.031799 & 34 & 3 & $C_s$ & 0.17556 & 0.17213 & 0.17399 \\
\hline
28 & 2 & $T_{d}$ & 0.12664 & 0.10097 & 0.10698 & 34 & 4 & $C_2$ & 0.17385 & 0.17383 & 0.17900 \\
\hline
30 & 1 & $D_{5h}$ & 0.22547 & 0.21665 & 0.23740 & 34 & 6 & $C_{3v}$ & 0.14952 & 0.15180 & 0.15330 \\
\hline
30 & 3 & $C_{2v}$ & 0.70608 & 0.72026 & 0.72119 & 36 & 1 & $C_2$ & 0.79026 & 0.81610 & 0.81657 \\
\hline
30 & 3 & $C_{2v}$ & 0.87899 & 0.89230 & 0.89357 & 36 & 3 & $C_1$ & 0.081562 & 0.076166 & 0.079149 \\
\hline
32 & 3 & $D_{3d}$ & 0.039925 & 0.035465 & 0.045702 & 36 & 3 & $C_1$ & 0.16074 & 0.15538 & 0.15864 \\
\hline
32 & 3 & $D_{3d}$ & 0.22246 & 0.21976 & 0.23086 & 36 & 4 & $C_s$ & 0.19945 & 0.19911 & 0.20050 \\
\hline
32 & 3 & $D_{3d}$ & 0.41451 & 0.43330 & 0.43509 & 36 & 4 & $C_s$ & 0.20320 & 0.20448 & 0.20625 \\
\hline
32 & 4 & $C_2$ & 0.23631 & 0.23543 & 0.24331 & 36 & 5 & $D_2$ & 0.11678 & 0.10602 & 0.11827 \\
\hline
32 & 5 & $D_{3h}$ & 0.84945 & 0.87440 & 0.87442 & 36 & 5 & $D_2$ & 0.68433 & 0.70262 & 0.70522 \\
\hline
32 & 5 & $D_{3h}$ & 0.85357 & 0.87843 & 0.87848 & 36 & 6 & $D_{2d}$ & 0.50607 & 0.51706 & 0.51979 \\
\end{tabular}
\label{table:2}
\end{center}
\end{table}

\begin{table}[h]
\begin{center}
\caption{Number of vertices $N$, isomer index no., spatial symmetry group, and magnetic field $h_{\chi}$ of the susceptibility discontinuity with respect to the saturation field $h_{sat}$ for classical spins.}
\begin{tabular}{c|c|c|c|c|c|c|c|c|c|c|c}
$N$ & no. & group & $h_{\chi} / h_{sat}$ & $N$ & no. & group & $h_{\chi} / h_{sat}$ & $N$ & no. & group & $h_{\chi} / h_{sat}$ \\
\hline
24 & 1 & $D_{6d}$ & 0.9381 & 32 & 4 & $C_2$ & 0.9617 & 36 & 5 & $D_2$ & 0.11935 \\
\hline
26 & 1 & $D_{3h}$ & 0.01142 & 32 & 5 & $D_{3h}$ & 0.8556 & 36 & 5 & $D_2$ & 0.9131 \\
\hline
26 & 1 & $D_{3h}$ & 0.1053 & 32 & 6 & $D_3$ & 0.07474 & 36 & 6 & $D_{2d}$ & 0.12156 \\
\hline
28 & 1 & $D_2$ & 0.05036 & 32 & 6 & $D_3$ & 0.1532 & 36 & 6 & $D_{2d}$ & 0.89230 \\
\hline
28 & 1 & $D_2$ & 0.9400 & 34 & 1 & $C_2$ & 0.9131 & 36 & 7 & $C_1$ & 0.93423 \\
\hline
28 & 2 & $T_{d}$ & 0.1632 & 34 & 2 & $C_s$ & 0.07904 & 36 & 8 & $C_s$ & 0.13052 \\
\hline
28 & 2 & $T_{d}$ & 0.1892 & 34 & 2 & $C_s$ & 0.9560 & 36 & 8 & $C_s$ & 0.99253 \\
\hline
28 & 2 & $T_{d}$ & 0.9863 & 34 & 2 & $C_s$ & 0.9939 & 36 & 9 & $C_{2v}$ & 0.97260 \\
\hline
30 & 1 & $D_{5h}$ & 0.8681 & 34 & 3 & $C_s$ & 0.7958 & 36 & 10 & $C_2$ & 0.02946 \\
\hline
30 & 2 & $C_{2v}$ & 0.1317 & 34 & 3 & $C_s$ & 0.9563 & 36 & 10 & $C_2$ & 0.07741 \\
\hline
30 & 2 & $C_{2v}$ & 0.1450 & 34 & 4 & $C_2$ & 0.01522 & 36 & 10 & $C_2$ & 0.97684 \\
\hline
30 & 2 & $C_{2v}$ & 0.7943 & 34 & 4 & $C_2$ & 0.9179 & 36 & 11 & $C_2$ & 0.02921 \\
\hline
30 & 2 & $C_{2v}$ & 0.9904 & 34 & 5 & $C_2$ & 0.9804 & 36 & 11 & $C_2$ & 0.07112 \\
\hline
30 & 3 & $C_{2v}$ & 0.9755 & 34 & 6 & $C_{3v}$ & 0.9976 & 36 & 11 & $C_2$ & 0.9534 \\
\hline
32 & 1 & $C_2$ & 0.01780 & 36 & 1 & $C_2$ & 0.84365 & 36 & 12 & $C_2$ & 0.9512 \\
\hline
32 & 1 & $C_2$ & 0.08741 & 36 & 2 & $D_2$ & 0.85592 & 36 & 13 & $D_{3h}$ & 0.06135 \\
\hline
32 & 1 & $C_2$ & 0.9025 & 36 & 3 & $C_1$ & 0.98972 & 36 & 13 & $D_{3h}$ & 0.8284 \\
\hline
32 & 2 & $D_2$ & 0.8708 & 36 & 4 & $C_s$ & 0.16333 & 36 & 15 & $D_{6h}$ & 0.9429 \\
\hline
32 & 3 & $D_{3d}$ & 0.8927 & 36 & 4 & $C_s$ & 0.81183 & & & \\
\hline
32 & 4 & $C_2$ & 0.05905 & 36 & 4 & $C_s$ & 0.9974 & & & \\
\end{tabular}
\label{table:3}
\end{center}
\end{table}


\begin{table}[h]
\begin{center}
\caption{Number of vertices $N$, isomer index no., spatial symmetry group, its corresponding number of operations, ground-state energy per spin $\frac{E_g}{N}$, full symmetry group multiplicity of the ground state mult., and saturation field $h_{sat}$ for $s=\frac{1}{2}$. The numbers have been calculated with double precision accuracy, but less significant digits are given where applicable for the sake of brevity.}
\begin{tabular}{c|c|c|c|c|c|c}
$N$ & no. & Symmetry & Number of & $\frac{E_g}{N}$ & mult. & $h_{sat}$ \\
 & & group & symmetry & & \\
 & & & operations & & \\
\hline
28 & 1 & $D_2$ & 4 & -0.49501 & 1 & 5.52469 \\
\hline
30 & 2 & $C_{2v}$ & 4 & -0.49633 & 1 & 5.55893 \\
\hline
30 & 3 & $C_{2v}$ & 4 & -0.49399 & 1 & 5.52183 \\
\hline
34 & 6 & $C_{3v}$ & 6 & -0.49667 & 2 & 5.58911 \\
\hline
36 & 15 & $D_{6h}$ & 24 & -0.50156 & 1 & 3 + $\sqrt{7}$ \\
\end{tabular}
\label{table:spinonehalf}
\end{center}
\end{table}

\begin{table}[h]
\begin{center}
\caption{Energy $E_g$, full symmetry group multiplicity mult., and irreducible representation irrep. of the lowest lying level in each $S$ sector for $s=\frac{1}{2}$. The irreducible representation notation follows Ref. \cite{Altmann94}. Each energy level corresponds to a multiplet of mult.$\times (2S+1)$ states. The numbers have been calculated with double precision accuracy, but less significant digits are given where applicable for the sake of brevity.}
\begin{tabular}{c|c|c|c|c|c|c|c|c|c|c|c}
\multicolumn{4}{c} {N=28, no. 1} & \multicolumn{4}{c} {N=30, no. 2} & \multicolumn{4}{c} {N=30, no. 3} \\
$S$ & $E_g$ & mult. & irrep. & $S$ & $E_g$ & mult. & irrep. & $S$ & $E_g$ & mult. & irrep. \\
\hline
 0 & -13.86022 & 1 & $A$ & 0 & -14.88985 & 1 & $B_2$ & 0 & -14.81980 & 1 & $B_1$ \\
\hline
 1 & -13.60531 & 1 & $B_1$ & 1 & -14.68682 & 1 & $A_2$ & 1 & -14.70370 & 1 & $A_1$ \\
\hline
 2 & -13.15371 & 1 & $A$ & 2 & -14.25828 & 1 & $B_2$ & 2 & -14.40830 & 1 & $B_1$ \\
\hline
 3 & -12.40902 & 1 & $B_2$ & 3 & -13.54806 & 1 & $A_1$ & 3 & -13.61215 & 1 & $B_1$ \\
\hline
 4 & -11.38331 & 1 & $A$ & 4 & -12.56921 & 1 & $B_2$ & 4 & -12.62424 & 1 & $A_1$ \\
\hline
 5 & -10.09449 & 1 & $B_2$ & 5 & -11.32412 & 1 & $A_1$ & 5 & -11.40301 & 1 & $A_1$ \\
\hline
 6 & -8.38100 & 1 & $A$ & 6 & -9.84464 & 1 & $B_2$ & 6 & -9.84064 & 1 & $B_1$ \\
\hline
 7 & -6.50948 & 1 & $B_2$ & 7 & -8.00704 & 1 & $B_2$ & 7 & -8.08381 & 1 & $A_1$ \\
\hline
 8 & -4.52581 & 1 & $A$ & 8 & -6.10058 & 1 & $B_2$ & 8 & -6.08468 & 1 & $B_1$ \\
\hline
 9 & -2.33269 & 1 & $B_2$ & 9 & -3.97609 & 1 & $A_1$ & 9 & -3.97409 & 1 & $A_1$ \\
\hline
10 & -0.017341 & 1 & $A$ & 10 & -1.75459 & 1 & $B_1$ & 10 & -1.75850 & 1 & $B_1$ \\
\hline
11 & 2.44260 & 1 & $B_1$ & 11 & 0.61476 & 1 & $A_1$ & 11 & 0.61462 & 1 & $A_2$ \\
\hline
12 & 5.00590 & 1 & $A$ & 12 & 3.11675 & 1 & $B_2$ & 12 & 3.12252 & 1 & $B_1$ \\
\hline
13 & 7.73765 & 1 & $B_1$ & 13 & 5.73337 & 1 & $A_2$ & 13 & 5.75602 & 1 & $A_1$ \\
\hline
14 & $\frac{21}{2}$ & 1 & $A$ & 14 & 8.47054 & 1 & $B_1$ & 14 & 8.48908 & 1 & $B_2$ \\
\hline
 & & & & & $\frac{45}{4}$ & 1 & $A_1$ & 15 & $\frac{45}{4}$ & 1 & $A_1$ \\
\end{tabular}
\begin{tabular}{c|c|c|c|c|c|c|c}
\multicolumn{4}{c} {N=34, no. 6} & \multicolumn{4}{c} {N=36, no. 15} \\
$S$ & $E_g$ & mult. & irrep. & $S$ & $E_g$ & mult. & irrep. \\
\hline
 0 & -16.88695 & 2 & $E$ & 0 & -18.05633 & 1 & $A_{1g}$ \\
\hline
 1 & -16.81690 & 2 & $E$ & 1 & -17.94985 & 1 & $B_{2g}$ \\
\hline
 2 & -16.59920 & 1 & $A_2$ & 2 & -17.56621 & 1 & $B_{1u}$ \\
\hline
 3 & -15.92485 & 1 & $A_1$ & 3 & -16.98683 & 1 & $A_{2u}$ \\
\hline
 4 & -14.95633 & 1 & $A_2$ & 4 & -16.17820 & 1 & $A_{1g}$ \\
\hline
 5 & -13.81712 & 1 & $A_1$ & 5 & -15.12249 & 1 & $A_{2u}$ \\
\hline
 6 & -12.50564 & 1 & $A_1$ & 6 & -13.86987 & 1 & $A_{1g}$ \\
\hline
 7 & -10.96011 & 1 & $A_1$ & 7 & -12.38258 & 1 & $A_{2u}$ \\
\hline
 8 & -9.14849 & 1 & $A_2$ & 8 & -10.74704 & 1 & $A_{1g}$ \\
\hline
 9 & -7.14688 & 1 & $A_2$ & 9 & -8.75446 & 1 & $B_{1u}$ \\
\hline
10 & -5.03424 & 1 & $A_1$ & 10 & -6.68639 & 1 & $A_{1g}$ \\
\hline
11 & -2.81088 & 1 & $A_2$ & 11 & -4.56438 & 1 & $A_{2u}$ \\
\hline
12 & -0.48815 & 1 & $A_2$ & 12 & -2.30549 & 1 & $A_{1g}$ \\
\hline
13 & 1.99045 & 1 & $A_1$ & 13 & 0.10921 & 1 & $A_{2u}$ \\
\hline
14 & 4.55327 & 1 & $A_1$ & 14 & 2.65495 & 1 & $A_{1g}$ \\
\hline
15 & 7.17825 & 1 & $A_2$ & 15 & 5.25354 & 1 & $A_{2u}$ \\
\hline
16 & 9.95545 & 1 & $A_1$ & 16 & 7.91309 & 1 & $A_{1g}$ \\
\hline
17 & $\frac{51}{4}$ & 1 & $A_1$ & 17 & 10.67712 & 1 & $A_{2u}$ \\
\hline
18 & & & & 18 & $\frac{27}{2}$ & 1 & $A_{1g}$ \\
\end{tabular}
\label{table:spinonehalfenergies}
\end{center}
\end{table}

\end{document}